\documentclass[11pt]{article}
\usepackage{moriond,epsfig}
\usepackage{amssymb,amsmath}

\def\Journal#1#2#3#4{{#1} {\bf #2}, #3 (#4)}

\def\NCA{\em Nuovo Cimento}

\def\PLB{{\em Phys. Lett.}  B}
\def\PRL{\em Phys. Rev. Lett.}
\def\PRD{{\em Phys. Rev.} D}

\def\JHEP{\em JHEP}
\def\AJS{\em Astrophys. J. Suppl.}

\def\be{\begin{equation}}
\def\ee{\end{equation}}
\def\bea{\begin{eqnarray}}
\def\eea{\end{eqnarray}}

\def \snu  { {\tilde{\nu}}}

\newcommand{\lviol}{$\not\!{\rm L}~$}

\begin{document}
\vspace*{4cm}
\title{SNEUTRINO COLD DARK MATTER IN EXTENDED MSSM MODELS}

\author{ C. ARINA }

\address{Department of Theoretical Physics, Universit\`a di Torino, 1 Via Pietro Giuria,\\
10125 Torino, Italy}

\maketitle\abstracts{A thorough analysis of sneutrinos as dark matter candidates is performed, in different classes of supersymmetric models, as is typically done for the neutralino dark matter. First in the Minimal Supersymmetric Standard Model, sneutrinos are marginally compatible with existing experimental bounds, including direct detection, provided they compose a subdominant component of dark matter. Then supersymmetric models with the inclusion of right-handed fields and lepton-number violating terms are presented. These models are perfectly viable: they predict sneutrinos which are compatible with the current direct detection sensitivities.}

\section{Sneutrino in the Minimal Standard Supersymmetric Model}\label{sec:mssm}
\begin{figure}[t]
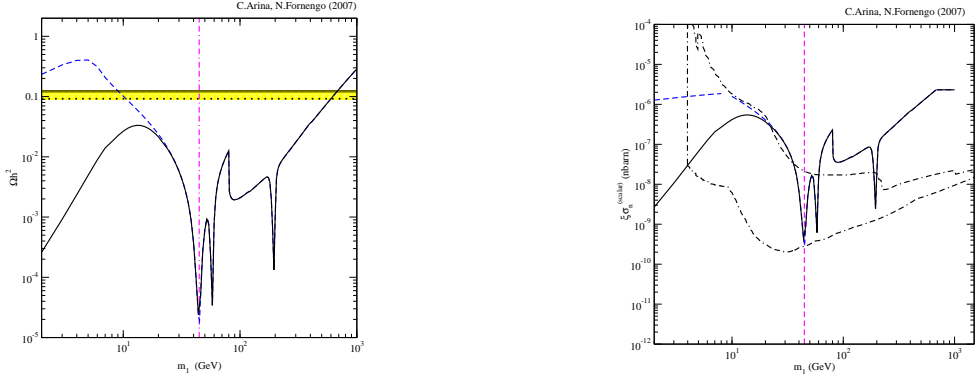

\hspace{-1cm}
\begin{minipage}[t]{0.5\textwidth} 
\centering
\psfig{figure=./figures/relic_mssm.eps,width=0.6\columnwidth}
\end{minipage}
\begin{minipage}[t]{0.5\textwidth}
\centering
\psfig{figure=./figures/direct_dama_mssm.eps,width=0.6\columnwidth}
\end{minipage}
\vspace*{-0.4cm}
\caption{Relic abundance $\Omega h^2$ (left) and sneutrino-nucleon scattering cross--section $\xi\sigma^{\rm scalar}_{\rm (nucleon)}$ (right) vs. $m_1$.}
\label{fig:mssm}
\end{figure}

\vspace*{-0.2cm}
We wish to reconsider in a consistent way sneutrino as a cold relic from the early
Universe and study its phenomenology relevant both for Cosmology and for relic--particle detection in low--energy supersymmetric extensions of the Standard Model, which does not (necessarily) invoke mSUGRA relations. We first review the phenomenology of the sneutrino as Cold Dark Matter (CDM) candidate in the case of the Minimal Supersymmetric extension of the Standard Model (MSSM). This model, not very appealing for sneutrino CDM and actually already almost excluded by direct detection searches sets the basis for the extended models described in the next section. In the MSSM, sneutrinos are the scalar partners of the left--handed neutrinos and are described by the usual superpotential and soft breaking terms, leading to the mass-term $m_{1} = \left[ m_{L}^{2} + \frac{1}{2} m_{Z}^{2} \cos 2\beta \right]$, where $m_L$ is the soft--mass for the left--handed SU(2) doublet $\tilde L$, $\beta$ is defined as usual from the relation $\tan\beta = v_{2}/v_{1}$, where $v_{2}$ is the
vacuum expectation value of the neutral component of the $H^{2}$ Higgs field and $m_{Z}$ is the $Z$--boson
mass. First of all, we have calculated the sneutrino relic abundance, by taking into account all the relevant annihilation channels and
co--annihilation processes which may arise when the sleptons are close in mass to the sneutrinos, as described in Ref.~\cite{af}. In this minimal MSSM models, the three sneutrinos are degenerate in mass: they therefore must be considered jointly
in the calculation of the relevant processes. An example of sneutrino relic abundance $\Omega h^2$ for the minimal MSSM is plotted in Fig. \ref{fig:mssm} as a function of the sneutrino mass $m_1$, where the vertical line denotes the lower bound on the sneutrino mass coming from the invisible $Z$--width and the solid (dashed) curves  refer to models with (without) gaugino universality. The sneutrino relic abundance is typically very small~\cite{fos}, much lower than the cosmological range for cold dark matter derived by the WMAP analysis~\cite{wmap}, which is $0.092\le\Omega_{\rm CDM } h^2 \le 0.124$ (yellow band, in between the horizontal black and dashed lines). For all the mass range from the experimental lower bound of about $m_{\rm Z}/2$ up to 600--700 GeV sneutrinos as the LSP are cosmologically acceptable but they are typically underabundant. This means that sneutrinos in the minimal version of MSSM are not good dark matter candidates, except for masses in a  narrow range which we determine to be 600--700 GeV. Dark matter direct search, which relies on the possibility to detect the recoil energy of a nucleus due to
the elastic scattering of the dark matter particle off the nucleus of a low--background detector, is known to
be a strong experimental constraint for sneutrino dark matter. The dependence of the direct detection rate on the DM particle rests into the particle mass and the scattering cross section. For sneutrinos, see Ref.~\cite{af} for details and references, coherent scattering arises due to $Z$ and Higgs exchange diagrams in the $t$--channel, therefore the
relevant cross section on nucleus is given by $ \sigma_{\cal N}=\sigma_{\cal N}^{Z}+\sigma_{\cal N}^{h,H}$. Comparisons with experimental results are more easily and typically performed by using the cross section
on a single nucleon $\xi \sigma^{\rm (scalar)}_{\rm nucleon}$. We have to consider that, whenever the dark matter particle is subdominant in the Universe, also its local density $\rho_{0}$ in the Galaxy is very likely reduced with respect to the total dark matter density. This means that the dominant component of dark matter is not the sneutrino, but still sneutrinos form a small amount of dark matter
and may be eventually detectable. In this case we rescale the local sneutrino abundance by means of the
usual factor $\xi = \min (1, \Omega h^{2}/0.092)$. When compared with the DAMA/NaI annual modulation region~\cite{rb} (dashed dotted curve) in Fig. \ref{fig:mssm} we see that direct detection is indeed a strong constraint on sneutrino dark matter in the minimal MSSM ~\cite{fos}, 
but some very specific configurations are still viable and could explain the annual modulation effect. 
\vspace*{-0.4cm}

\section{Non minimal supersymmetric models}\label{sec:ext}
\vspace*{-0.2cm}
The models which we will be considering are natural and direct extensions of the MSSM which incorporate at the same time the new physics required to explain two basic problems of astro--particle physics: the origin of neutrino masses and the nature of dark matter. 
The first class of models (LR models), enlarge the neutrino/sneutrino sector by the inclusion of sterile right--handed superfields 
$\hat N^{I}$ \cite{ah,gh}.
The relevant terms in the superpotential and in the soft breaking potential are:
\vspace*{-0.15cm}
\begin{eqnarray}\label{w_lr}
W & = & \epsilon_{ij} (\mu \hat H^{1}_{i} \hat H^{2}_{j} - Y_{l}^{IJ} \hat H^{1}_{i} \hat L^{I}_{j} \hat R^{J}
+ Y_{\nu}^{IJ} \hat H^{2}_{i} \hat L^{I}_{j} \hat N^{J} )\nonumber\\
V_{\rm soft} & = & (M_{L}^{2})^{IJ} \, \tilde L_{i}^{I \ast} \tilde L_{i}^{J} + 
(M_{N}^{2})^{IJ} \, \tilde N^{I \ast} \tilde N^{J} - 
 [\epsilon_{ij}(\Lambda_{l}^{IJ} H^{1}_{i} \tilde L^{I}_{j} \tilde R^{J} + 
\Lambda_{\nu}^{IJ} H^{2}_{i} \tilde L^{I}_{j} \tilde N^{J})  + \mbox{h.c.}]
\end{eqnarray}
where $M_{N}^{2}$, $\Lambda_{\nu}^{IJ}$, $M_{L}^{2}$, $\Lambda_{l}^{IJ}$ and $Y_{\nu}^{IJ}$ are matrices, which we choose real and diagonal. The Dirac mass of neutrinos are obtained $m_D^{I} = v_{2}Y_{\nu}^{II}$ from $Y_{\nu}^{IJ}$. The parameter which mixes the left-- and right--handed sneutrino fields may naturally be of the order of the other entries of the matrix, and induce a sizeable mixing of the lightest sneutrino in terms of left--handed and right--handed fields. The lightest mass eigenstate is therefore $\snu_1 = -\sin\theta\;\snu_L + \cos\theta\;\tilde{N}$, where $\theta$ is the LR mixing angle. Sizeable mixings reduce the coupling to the $Z$--boson, which couples only to left--handed fields, and therefore have relevant impact on all the sneutrino phenomenology: the lightest sneutrino may be lighter than $m_{Z}/2$ and the $\snu_{1}$ annihilation and scattering cross sections which involve $Z$ exchange are reduced, see Ref~\cite{af} for details. A supersymmetric model which can accommodate a Majorana mass--term for neutrinos and explain the 
observed neutrino mass pattern, may be built by adding to the minimal MSSM right--handed fields $\tilde N^{I}$
and allowing for lepton number violating (\lviol) terms (Majorana models). The most general form of the superpotential~\cite{ah,gh} and of the soft breaking potential~\cite{dhr}, which accomplishes this conditions is:
\begin{eqnarray}\label{w_maj}
 W & = & \epsilon_{ij} (\mu \hat H^{1}_{i} \hat H^{2}_{j} - Y_{l}^{IJ} \hat H^{1}_{i} \hat L^{I}_{j} \hat R^{J}
+ Y_{\nu}^{IJ} \hat H^{2}_{i} \hat L^{I}_{j} \hat N^{J} ) + \frac{1}{2} M^{IJ} \hat N^I \hat N^J\nonumber\\
V_{\rm soft} & = & (M_{L}^{2})^{IJ} \, \tilde L_{i}^{I \ast} \tilde L_{i}^{J} + 
(M_{N}^{2})^{IJ} \, \tilde N^{I \ast} \tilde N^{J} - \nonumber\\
& & [(m^2_B)^{IJ}\tilde{N}^I\tilde{N}^J+\epsilon_{ij}(\Lambda_{l}^{IJ} H^{1}_{i} \tilde L^{I}_{j} \tilde R^{J} + 
\Lambda_{\nu}^{IJ} H^{2}_{i} \tilde L^{I}_{j} \tilde N^{J})  + \mbox{h.c.}]
\end{eqnarray}
where we again use the same assumptions of diagonality in flavour space for all the matrices  as we already did before. For the \lviol parameters we therefore assume: $M^{IJ}=M\;\delta^{IJ}$, in order to reduce the number of free parameters. The Dirac mass of the neutrinos is obtained as $m_D^{I} = v_{2}Y_{\nu}^{II}$, while $M$ represent a Majorana mass--term for neutrinos. The Dirac--mass parameter is derived by the condition that the neutrino mass is determined by the see-saw mechanism: $m^{I}_{\nu}=m^{I}_{D}/M^{2}$. Sneutrinos now are a superpositions of two complex fields: the left--handed field $\nu_{L}$ and the right--handed field $\tilde N$. Since we introduced \lviol terms, it is convenient to work in a basis of CP eigenstates, where the $\snu-Z$ coupling is off diagonal. The non-diagonal nature of the $Z$--coupling leads to important consequences: first of all annihilation processes through Z channel become co--annihilation processes, thus reducing the annihilation cross sections; moreover the elastic scattering off nucleon becomes an inelastic scattering via $t$--channel Z exchange and under certain kinematics condition may be suppressed, leading to a lower value of the direct detection rate of the sneutrino dark matter. The lightest state, which is our dark matter candidate, may now exhibit the non--diagonal nature of the $Z$--coupling with respect of the CP eigenstates, $\snu_-, \snu_+$, which are also mass eigenstates, and a mixing with the right--handed field $\tilde N$: $\snu_i= Z_{i1}\snu_{+}+Z_{i2}\tilde{N}_+ +Z_{i3}\snu_{-}+Z_{i4}\tilde{N}_{-}$ with $i=1,2,3,4$.

\begin{figure}[h]
\hspace{-1.5cm}
\vspace*{-0.5cm}
\begin{minipage}[t]{0.5\textwidth}
\centering
\psfig{figure=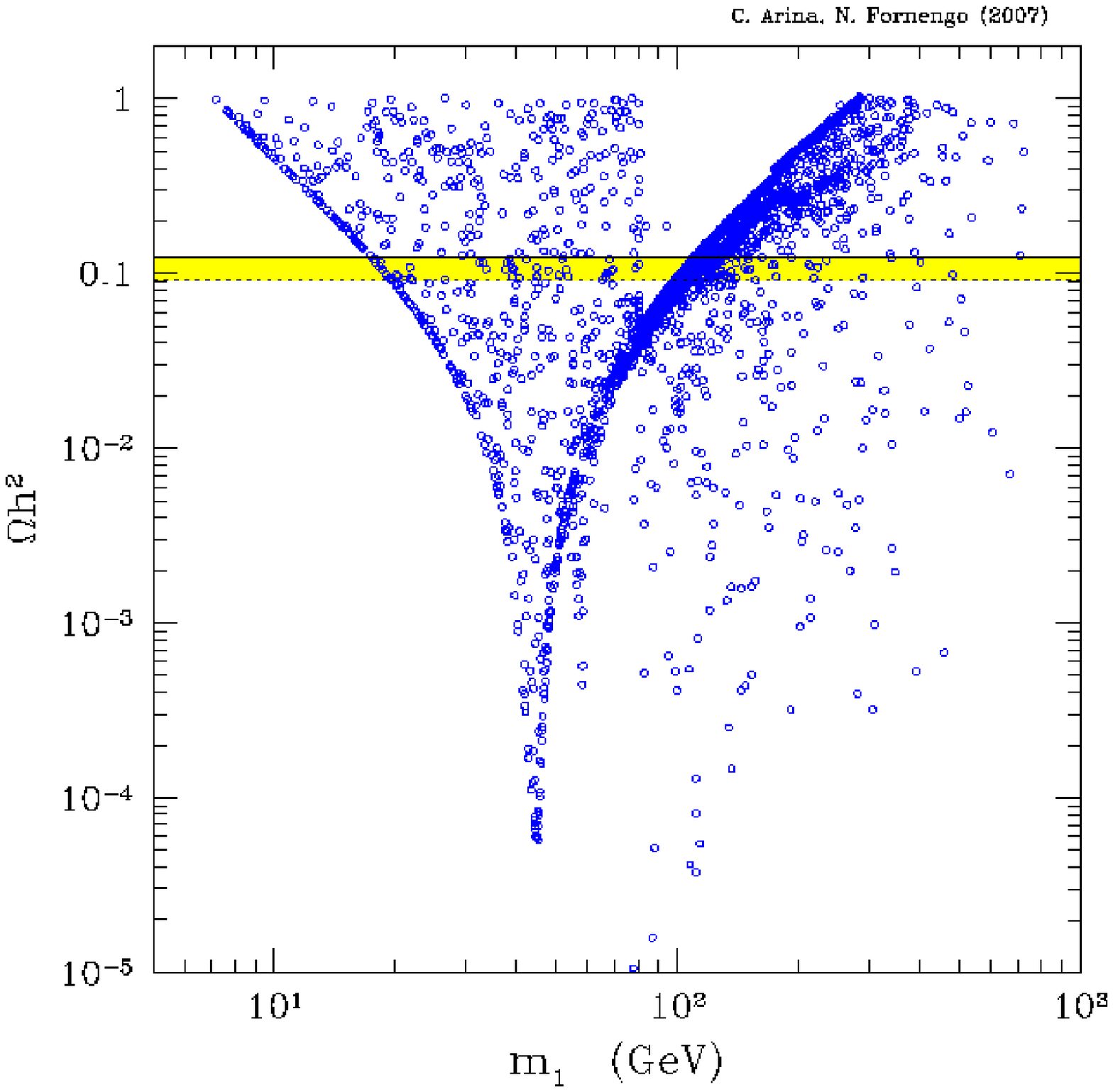,width=0.8\columnwidth}\\
\vspace*{-0.8cm}
\psfig{figure=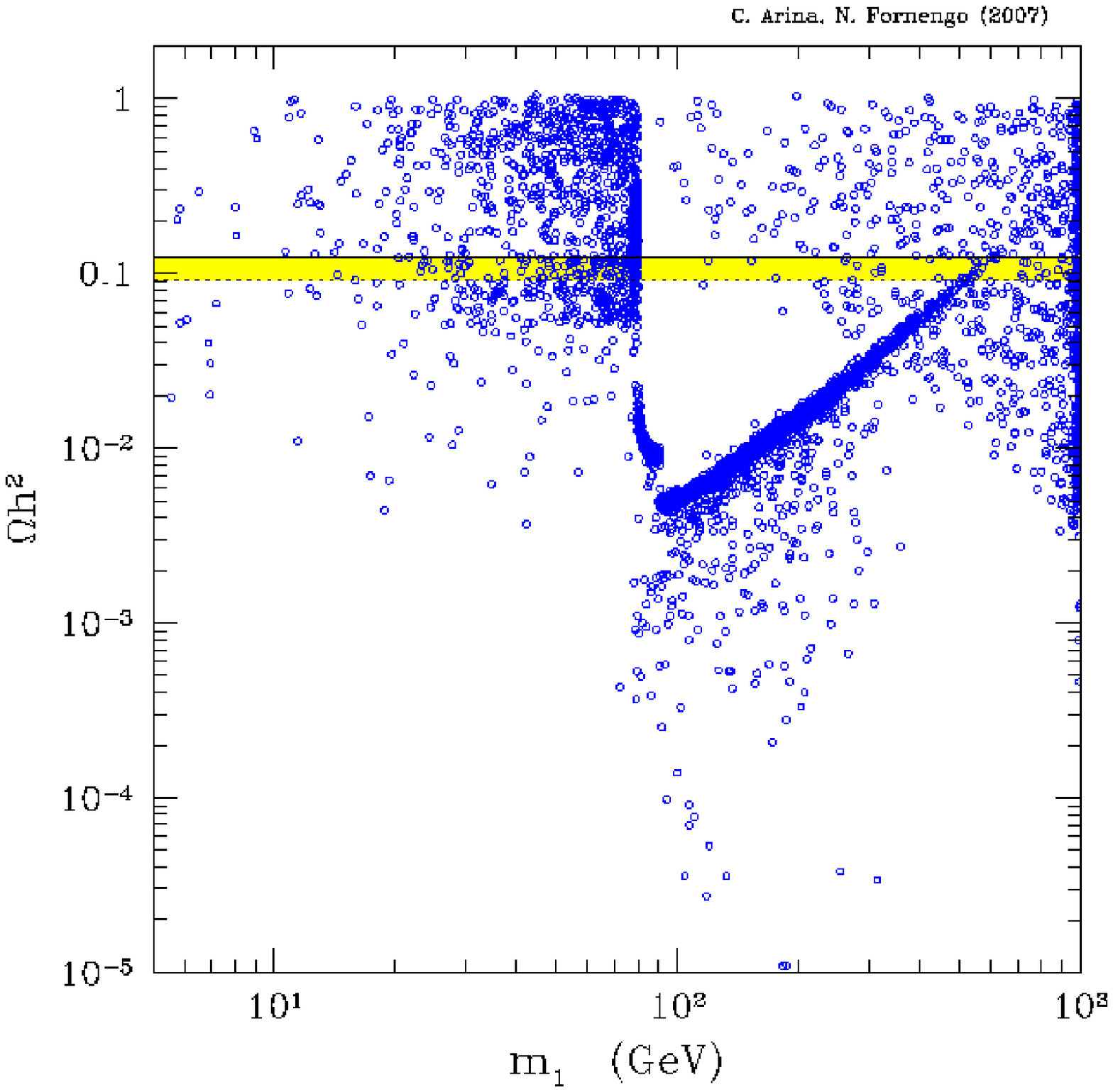,width=0.8\columnwidth}
\end{minipage}
\begin{minipage}[t]{0.5\textwidth}
\centering
\psfig{figure=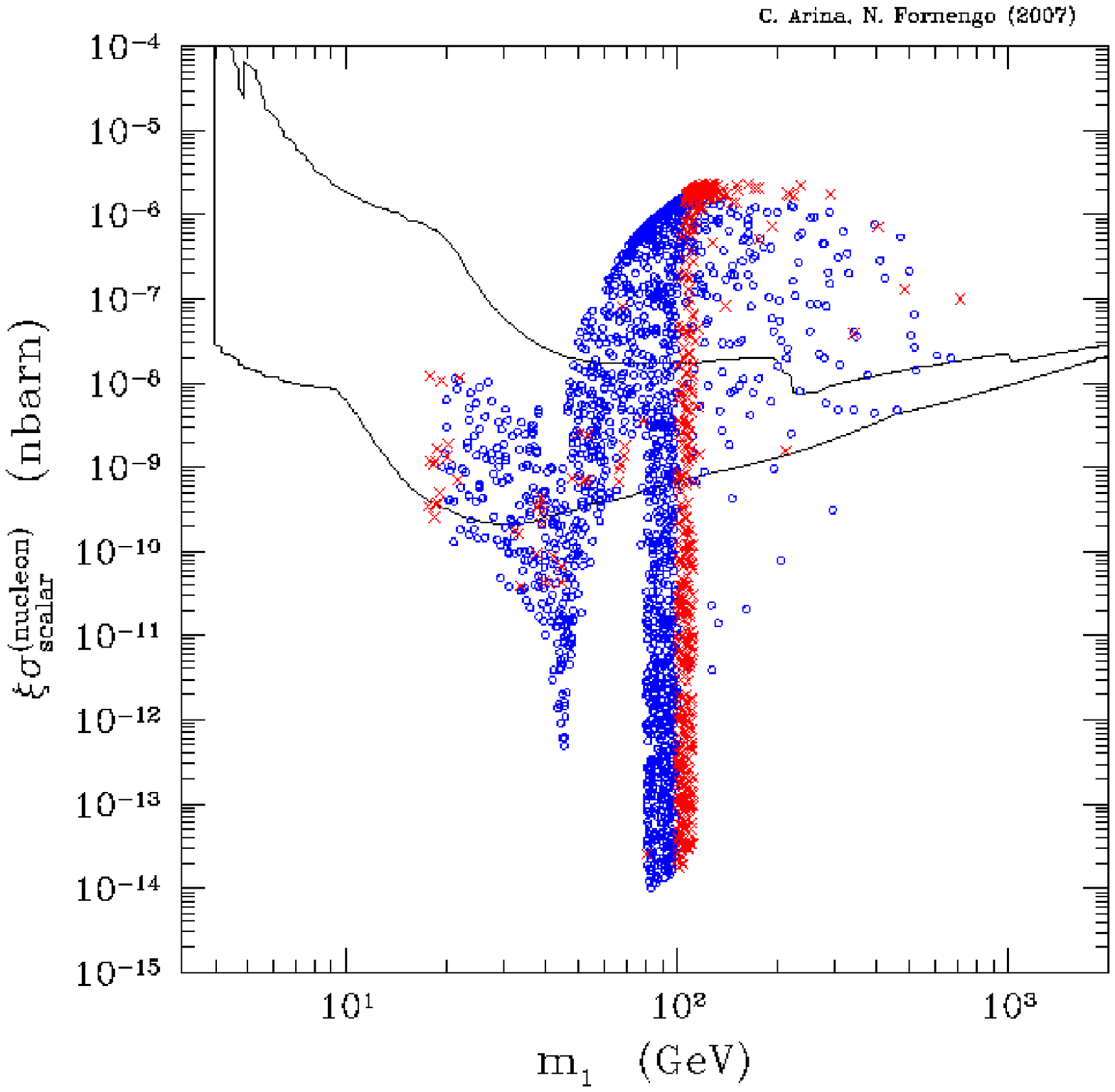,width=0.8\columnwidth}\\
\vspace*{-0.8cm}
\psfig{figure=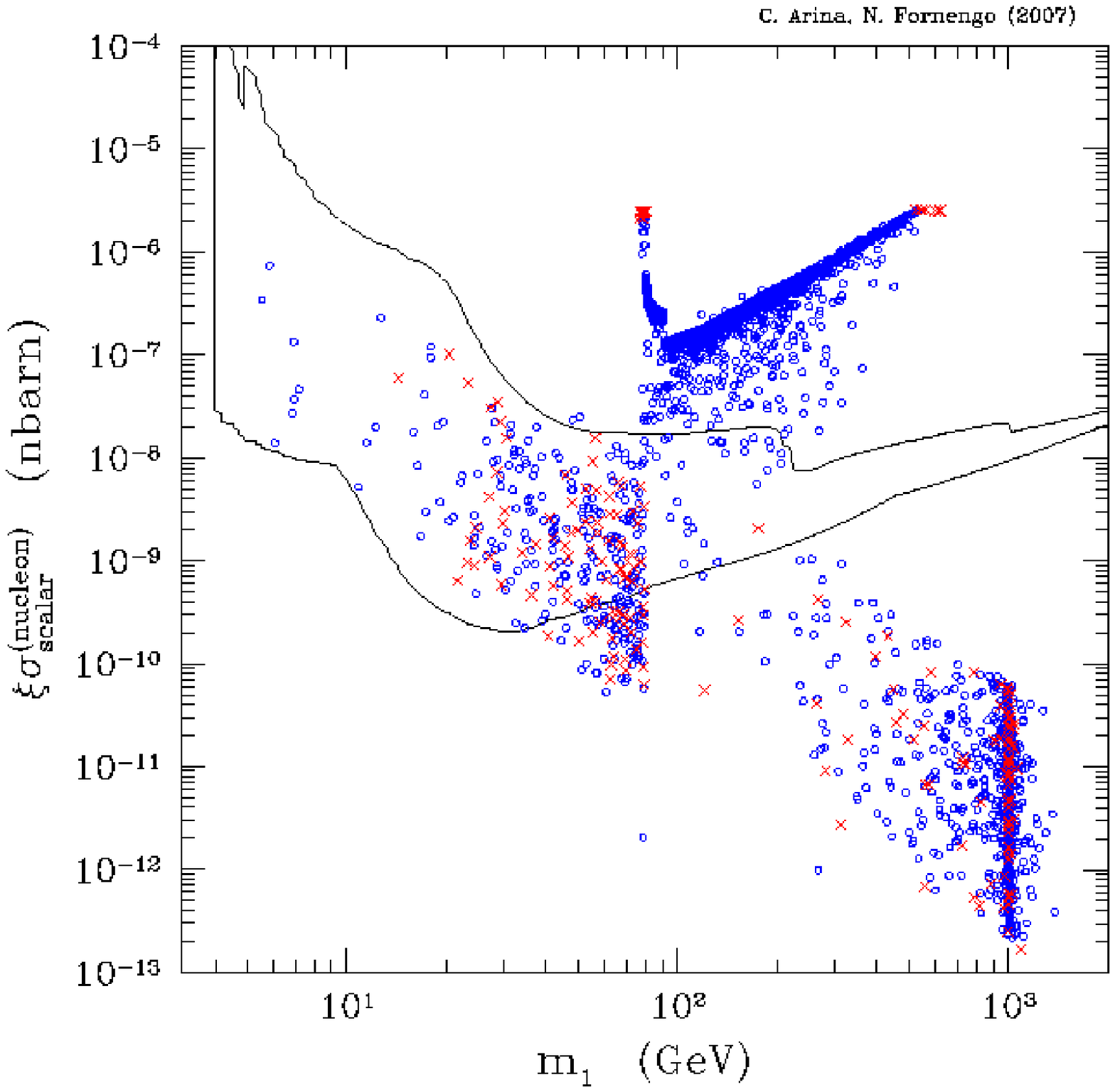,width=0.8\columnwidth}
\end{minipage}
\caption{Relic abundance $\Omega h^2$ (left) and sneutrino-nucleon scattering cross--section $\xi\sigma^{\rm scalar}_{\rm (nucleon)}$ (right) versus the sneutrino mass $m_1$ for LR models (top) and for low--scale Majorana models ($M=1$ TeV) (bottom).}
\label{fig:ext_mssm}
\end{figure}

In LR models sneutrinos may represent the dominant
dark matter component for a wide mass range. The most relevant new feature
is that for the full supersymmetric scan, the mass range allowed by the cosmological constraints is
enlarged up to 800 GeV, and all the mass interval above the $Z$--pole may lead to strongly subdominant
sneutrinos. From Fig. \ref{fig:ext_mssm}
we can conclude that after all experimental and theoretical constraints are imposed, sneutrino dark matter is perfectly viable, both as a dominant and
as a subdominant component, for the whole mass range $15~\mbox{GeV} \lesssim m_1 \lesssim 800~\mbox{GeV}$.
The lower limit of 15 GeV represents therefore a cosmological bound on the sneutrino mass in LR models,
under the assumption of $R$--parity conservation. The sneutrino--nucleon cross section is shown in the right top in Fig. \ref{fig:ext_mssm}. Only points which are
accepted by the cosmological constraint are shown. We see that the presence of the mixing with the
right--handed $\tilde N$ fields opens up the possibility to have viable sneutrino cold dark matter.
A fraction of the configurations are excluded by direct detection, but now, contrary to the minimal MSSM 
case, a large portion of the supersymmetric parameter space is compatible with the direct detection bound,
both for cosmologically dominant ([red] crosses) and subdominant ([blue] points) sneutrinos. 
The occurrence of sneutrinos which are not in conflict with direct detection limits and, at the same 
time, are the dominant dark matter component, is a very interesting feature of this class of models. The relic abundance of the Majorana models at low--mass scale is shown on the left bottom in Fig. \ref{fig:ext_mssm}. It is remarkable
that in the whole mass range from 5 GeV to 1 TeV sneutrinos can explain the required amount of CDM in the Universe. For the same mass range, sneutrinos may as well be a subdominant component. Direct detection is shown on the right bottom in Fig. \ref{fig:ext_mssm}. We see that three different populations
arise: configurations on the upper right are clearly excluded by direct detection searches.
Most of them refer to subdominant sneutrinos. Configurations on the lower right part of the
plot are allowed but well below current direct detection sensitivity. Configurations
on the center and left part of the plot all fall inside the current sensitivity region: a large fraction are cosmologically dominant and could explain the annual modulation effect observed by  the DAMA/NaI
experiment. 

\vspace*{-0.3cm}
\section*{Acknowledgments}
\vspace*{-0.2cm}
I acknowledge the European ``Marie Curie`` Programme for the support grant.
\vspace*{-0.4cm}
\section*{References}
\vspace*{-0.2cm}
\bibliographystyle{unsrt}

\end{document}